\documentclass[12pt,preprint]{aastex}
\usepackage{natbib}
\usepackage[normalem]{ulem}
\usepackage{rotating}

\begin{document}

\title{Interpreting the Spin-down Evolutions of Isolated Neutron Stars with Hall Effects}

\author{Yi Xie\altaffilmark{1,2} and Shuang-Nan Zhang\altaffilmark{1,3,4}}
\altaffiltext{1}{National Astronomical Observatories, Chinese
Academy Of Sciences, Beijing 100012, China}
\altaffiltext{2}{University of Chinese Academy of Sciences, Beijing
100049, China} \altaffiltext{3}{Laboratory for Particle
Astrophysics, Institute of High Energy Physics, Beijing 100049,
China} \altaffiltext{4}{Physics Department, University of Alabama in
Huntsville, Huntsville, AL 35899, USA}

\begin{abstract}

The observed long-term spin-down evolution of isolated radio pulsars
cannot be explained by the standard magnetic dipole radiation with a
constant braking torque. However, how and why the torque varies still
remains controversial, which is a major issue in
understanding neutron stars. Many pulsars have been observed with
significant long-term changes of their spin-down rates modulated by
quasi-periodic oscillations. Applying the phenomenological model of
pulsar timing noise we developed recently to the observed precise
pulsar timing data of isolated neutron stars, here we show that the observed long-term evolutions of their
spin-down rates and quasi-periodic modulations can be explained by Hall effects in their crusts. Therefore
the evolution of their crustal magnetic fields, rather than that in their cores, dominates the observed long term spin-down evolution of these young pulsars. Understanding of the nature of pulsar timing noise not only
reveals the interior physics of neutron stars, but also allows
physical modeling of pulsar spin-down and thus improves the
sensitivity of gravitational wave detections with pulsars.
\end{abstract}


\section{Introduction}

Pulsars are very stable natural clocks with observed steady pulses.
However, many pulsars exhibit significant timing irregularities,
i.e., unpredicted arrival times of pulses. Hobbs et al. (2010,
hereafter H2010) carried out so far the most extensive study of the
long-term timing irregularities of 366 pulsars, and ruled out some
timing noise models in terms of observational imperfections, random
walks (Boynton 1972; Alpar et al. 1986; Cheng 1987a, 1987b), and
planetary companions (Cordes \& Shannon 2008). Lyne et al. (2010,
hereafter L2010) found that timing behaviors often result from two
different spin-down rates, and pulsars switch abruptly between these
states, often quasi-periodically, leading to the observed spin-down
patterns. The deviations of pulsars' spin frequency $\dot\nu(t)$
from its long-term trend are correlated with changes in the pulse
shapes (see Fig. 3 and 4 of L2010), and therefore are magnetospheric
in origin. By modeling the observed precise timing data of pulsars,
we will show that the long-term linear change of $\dot\nu$ is
consistent with the timescale of Hall drift, and the oscillatory
structure of $\dot\nu$ and the changes in pulse shapes can be
produced by the Hall waves in the crust of a neutron star (NS).
Consequently the mechanism of magnetic field evolution and the
origin of the timing noise of pulsars are better understood, which
may improve the sensitivity of gravitational wave detections with
pulsars.

\section{Phenomenological model and observational test}

Assuming the pure magnetic dipole radiation in vacuum as the braking
mechanism for a pulsar's spin-down (e.g. Lorimer \& Kramer 2004), we
have
\begin{equation}\label{braking law}
\dot\nu =-A B(t)^2 \nu^3,
\end{equation}
in which $A=8\pi^2R^6\sin\theta^2/3c^3I$ is a constant, $B(t)$ is
the strength of the dipole magnetic field at the surface of the NS,
$R~(\simeq10^6~{\rm cm})$, $I~(\simeq10^{45}~{\rm g~cm^2})$, and
$\theta~(\simeq\pi/2)$ are the radius, moment of inertia, and angle
of magnetic inclination, respectively. Previously, we constructed a
phenomenological model for $B(t)$ with a long-term evolution
modulated by short-term oscillations (Zhang \& Xie 2012a and 2012b,
hereafter Paper~I and Paper~II, respectively):
\begin{equation}\label{B evolution}
B(t)=B_{\rm L}(t)(1+\sum \kappa_i\sin(\phi_i+2\pi\frac{t}{T_i})),
\end{equation}
where $t$ is the pulsar's age, and $\kappa_i(\ll 1)$, $\phi_i$, $T_i$
are the amplitude, phase and period of the $i$-th oscillating
component, respectively. $B_{\rm L}(t)=B_0(t/t_0)^{-\alpha}$, in
which $B_0$ is the field strength at age $t_0$, and $\alpha$ is
the power law index. Then from Equation (\ref{B evolution}) and
taking only the dominating oscillating component, we obtained the
analytic approximation for $\ddot\nu$ (Paper~I):
\begin{equation}\label{vddot}
\ddot{\nu}_{\pm}\simeq -2\dot{\nu}(\alpha/t_{\rm age}\pm f),
\end{equation}
where $t_{\rm age}$ is the real age of the pulsar and $f\equiv 2\pi
\kappa/T$ for the dominating oscillating component.

It has been found that Equation~(\ref{vddot}) describes adequately
the statistical properties of the observed timing noises of radio
pulsars (Paper~I). Therefore for relatively young pulsars with
$t_{\rm age}<3\times 10^5$~yr, the first term in
Equation~(\ref{vddot}) dominates and we should have $\ddot{\nu}>0$
if $\alpha >0$. Considering that the characteristic ages ($\tau_{\rm
c}$) of young pulsars are normally several times larger than $t_{\rm
age}$, Equation~(\ref{vddot}) thus explains naturally the observed
$\ddot\nu >0$ for most young pulsars with $\tau_{\rm c} \lesssim
10^6$ yr. Similarly for much older pulsars, the second term in
Equation~(\ref{vddot}) dominates, in agreement with the
observational fact that the numbers of negative and positive
$\ddot\nu$ are almost equal for the old pulsars in the sample of
H2010. The confirmation of Equation~(\ref{vddot}) with observations
naturally suggests that the observed pulsars indeed have
evolutionary links, and the equation therefore provides details
about the dipole magnetic field evolution of a young pulsar into an
old pulsar.

However, Equation~(\ref{vddot}) has so far not been tested with the
observed evolutions of individual pulsars. Following the same
approach in Paper~I, we obtain
\begin{equation}\label{vdot}
\dot{\nu}\simeq \dot\nu_0(1+2\Sigma
\kappa_i(\sin(\phi_i+2\pi\frac{t}{T_i})-\sin \phi_i))+\ddot\nu_{\rm
L}(t-t_0),
\end{equation}
where $\dot\nu_0=\dot\nu(t_0)$, $\ddot\nu_{\rm
L}=-2\alpha\dot\nu_0/t_0$ describes the long-term monotonic
variation of $\dot\nu(t)$. Therefore Equation~(\ref{vdot}) can be
tested with the long-term monitoring observations of individual
pulsars. We can obtain an estimate of $\ddot\nu_{\rm L}$ by
linearly fitting the long-term monotonic variation of reported
$\dot\nu(t)$. Consequently we can also determine the time scale of
the long-term magnetic field evolution of each pulsar (see
Equation~(6) in Paper~I)
\begin{equation}\label{tau_B}
\tau_{B}\equiv \frac{B}{\dot B}=\frac{2\dot\nu_0\nu_0}{\ddot\nu_{\rm
L}\nu_0 -3\dot\nu_0^2},
\end{equation}
for comparisons with theoretical models of neutron star magnetic
field evolution. $\tau_{B}<0$ indicates magnetic field decrease and
vice versa.

\begin{figure}
\center
\includegraphics[width=12 cm,angle=0]{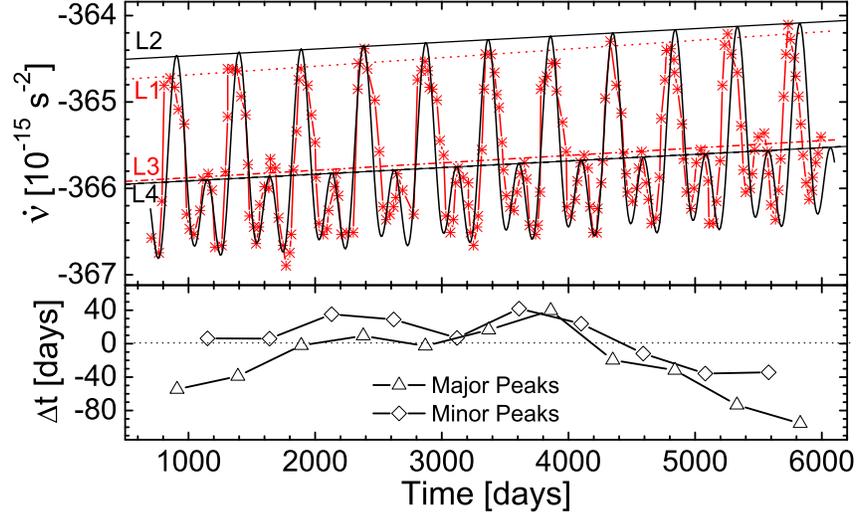}
\caption{\label{Fig:1} {\it Upper Panel}: $\dot\nu (t) $ for PSR
B1828$-$11 during the past 20 years. The reported data (taken from
L2010) are represented by red stars; and the solid black line is
calculated from Equation~(\ref{vdot}) with the following parameters:
$T_1=2T_2=500$~days, $\kappa_1=\kappa_2=10^{-3}$, $\phi_1=5.6 $,
$\phi_1=0.5$. The dotted, solid, dot-dashed and dashed lines are
linear fits for the major peak values of reported data and
analytical calculation, the minor peak values of reported data and
analytical calculation, respectively. {\it Lower Panel}: Time
differences between the peak positions of reported data and
analytical calculation. The differences for major peak positions and
minor peak positions are represented by triangles and diamonds,
respectively.}
\end{figure}

The sample of L2010 provides the precise histories of $\dot\nu$ for
seventeen pulsars and thus may be applied to test
Equation~(\ref{vdot}). Lyne et al. (2010) found that PSR B1828$-$11 clearly shows
a long-term evolution trend (noticed in L2010 as a linear increase of its $\dot\nu$), i.e. $\ddot\nu_{\rm L}>0$.
We show the comparison between the reported (taken
from L2010) and analytically calculated $\dot\nu(t)$ for the pulsar in Figure \ref{Fig:1}; all the parameters are listed in Table 1.
The one major difference is caused by the decrease of the
oscillation periods of the reported data after $\sim4000$~days,
which indicates that the oscillations are unlikely caused by
precession of the neutron star. Nevertheless, our model describes
the general trend of the reported data of B1828 quite well.

By linearly fitting $\dot\nu$ of all other pulsars, we find that most of them also exhibit long-term monotonic evolutions modulated by
short-term quasi-periodical oscillations. However, three pulsars in the sample, namely, B1822-09, B2035+36 and J2043+2740, exhibit much more erratic
short-term behaviors, which may significantly bias the fitting results; here we exclude these three pulsars from further study. We conclude that our phenomenological model can describe adequately the spin-down evolutions of fourteen pulsars in the sample of seventeen pulsars of L2010; all the observed and
derived parameters for the fourteen pulsars are listed in Table 1.

\section{Physical implications}

\subsection{Inclination evolution?}

We first investigate the alternative possibility that the long-term
linear change of $\dot\nu(t)$ is caused by the change of the
inclination angle of a NS, rather than the change of its magnetic
field strength. Observationally, the inclination angle change would
also produce a long-term change on the width of pulses. For a
circular beam, the change of its width can be given by the
geometrical relation
\begin{equation}\label{Width}
\Delta W=W_0-4\arcsin [\frac{(\sin ^2 \frac{\gamma}{2}-\sin
^2\frac{\beta +\Delta\theta}{2})}{\sin (\theta-\Delta\theta)
\sin(\theta+ \beta )}]^{1/2},
\end{equation}
where $\Delta\theta$ is the magnitude of the inclination angle
change that can be obtained from
$\sin^2\Delta\theta/\sin^2\theta\simeq\ddot \nu_{\rm L}\Delta t/\dot
\nu_0 \sim 0.0017$ for PSR B1828$-$11 ($\Delta t$ is the time span of the
observations), $\beta$ is the impact angle, $\gamma$ is the half
angular width of the radiating cone (Gil et al. 1984),
\begin{equation}\label{Beamwidth}
\gamma=2\arcsin[\sin^2\frac{W_0}{4}\sin\theta\sin(\theta+\beta)+\sin^2\frac{\beta}{2}]^{\frac{1}{2}},
\end{equation}
$W_0\sim 0.04 P$ is the initial pulse profile width of B1828$-$11
(L2010). Thus, $\Delta W$ only depends on $\theta$ and $\beta$.
There are two observational constraints for $\theta$ and $\beta$ for
B1828$-$11: (a) $\Delta W\simeq -0.28\pm 0.51$ ms (L2010); and (b)
its position angle (P.A.) gives $\sin\theta/\sin\beta\sim 5.35$
(Gould \& Lyne 1998).

\begin{figure}
\center
\includegraphics[width=10 cm,angle=0]{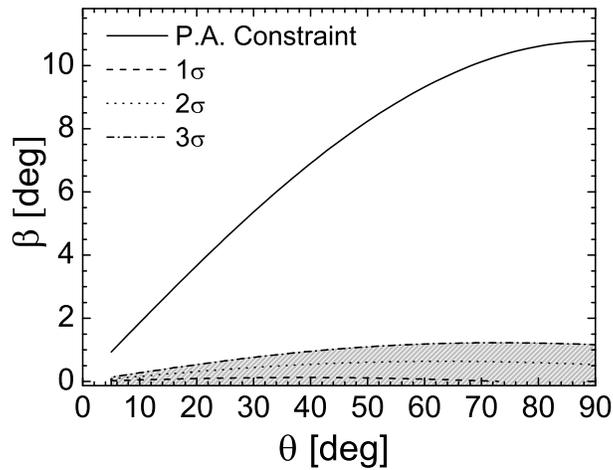}
\caption{\label{Fig:3} Comparison between the allowed area in the
$\theta-\beta$ plane from $\Delta W$ and that from the position
angle constraint for B1828$-$11. }
\end{figure}

We show the comparison of the two constraints on $\theta$ and
$\beta$ in Figure \ref{Fig:3}. One can see that condition (a)
requires $\beta\lesssim 1.2~{\rm deg}$, which is much smaller than
that from condition (b); thus the possibility of inclination angle
change is ruled out for B1828$-$11. It is therefore reasonable to assume that the long-term linear changes of $\dot\nu(t)$ for other pulsars in the sample are also not caused by the changes of their inclination angles, because the patterns of their spin-down evolutions are very similar to PSR B1828$-$11.

\subsection{Long-term magnetic field evolution}

Goldreich \& Reisenegger (1992) studied several avenues for magnetic
field decay in isolated NSs: ohmic decay, ambipolar diffusion, and
Hall drift. Depending on the strength of the magnetic fields, each
of these processes may dominate the evolution, and the ambipolar
diffusion is only important for magnetars. Ohmic decay occurs in
both the fluid core and solid crust of a NS (Sang \& Chanmugam 1987;
Urpin et al. 1994; Page et al. 2000). It is inversely proportional
to the electric conductivity and independent of the strength of
magnetic fields. If an electric current flows vertically in magnetic
fields, the fields exerts a transverse force on the moving charge
carriers. The resulting Hall drift of the carriers can transport
magnetic fields from the inner crust to the outer crust (Jones
1988). The Hall effect is non-dissipative and thus cannot be a
direct cause of magnetic field decay. However, it can enhance the
rate of ohmic dissipation, since only electrons are mobile in the
solid crust, and their Hall angle is large. Consequently the
evolution of magnetic fields resembles that of vorticity, and then
the fields undergo a turbulent cascade terminated by ohmic
dissipation at small scales (Goldreich \& Reisenegger 1992; Muslimov
1994; Biskamp \& M$\ddot{u}$ller 1999; Urpin \& Shalybkov 1999;
Rheinhardt \& Geppert 2002; Geppert \& Rheinhardt 2002; Pons \&
Geppert 2010; Geppert et al. 2013).

Cumming et al. (2004, hereafter C2004) found that, in isolated NSs
with relatively pure crusts, the Hall effect dominates over ohmic
decay after a time $t_{\rm switch}\simeq 10^4 B_{12}^{-3}~{\rm yr}$,
where $B_{12}$ is in units of $10^{12}$ G. This is consistent with
the fact that $\tau_{\rm B}$ is much shorter than the ohmic
timescale $\tau_{\rm ohm}\gtrsim 10^6$ yr. At lower densities of the
crust, the degenerated electrons contribute to the pressure, and the
Hall timescale ($\tau_{\rm Hall}\equiv B/|\dot B_{\rm Hall}|$, where
$\dot B_{\rm Hall}$ is the decay rate induced by Hall effect) is
(C2004)
\begin{equation}\label{Hallout}
\tau_{\rm Hall,~Outer}=\frac{5.7\times 10^4~{\rm
yr}}{B_{12}}\rho_{12}^{5/3}(\frac{Y_e}{0.25})^{11/3}(\frac{g_{14}}{2.45})^{-2},
\end{equation}
in which $\rho_{12}=\rho/10^{12}~{\rm g~cm^{-3}}$, $Y_e$ is the
number fraction of electrons, and $g_{14}$ is the local gravity,
assumed constant. At densities greater than neutron drip, the
neutrons dominate the pressure (C2004),
\begin{equation}\label{Hallin}
\tau_{\rm Hall,~Inner}=\frac{1.2\times 10^7~{\rm
yr}}{B_{12}}\rho_{14}^{7/3}Y_n^{10/3}(\frac{Y_e}{0.05})(\frac{f_n}{0.5})^2(\frac{g_{14}}{2.45})^{-2},
\end{equation}
where $Y_n$ is the number fraction of neutrons, and $f_n$ is the
factor that accounts for the interactions between neutrons.

\begin{figure}
\center
\includegraphics[width=18 cm,angle=0]{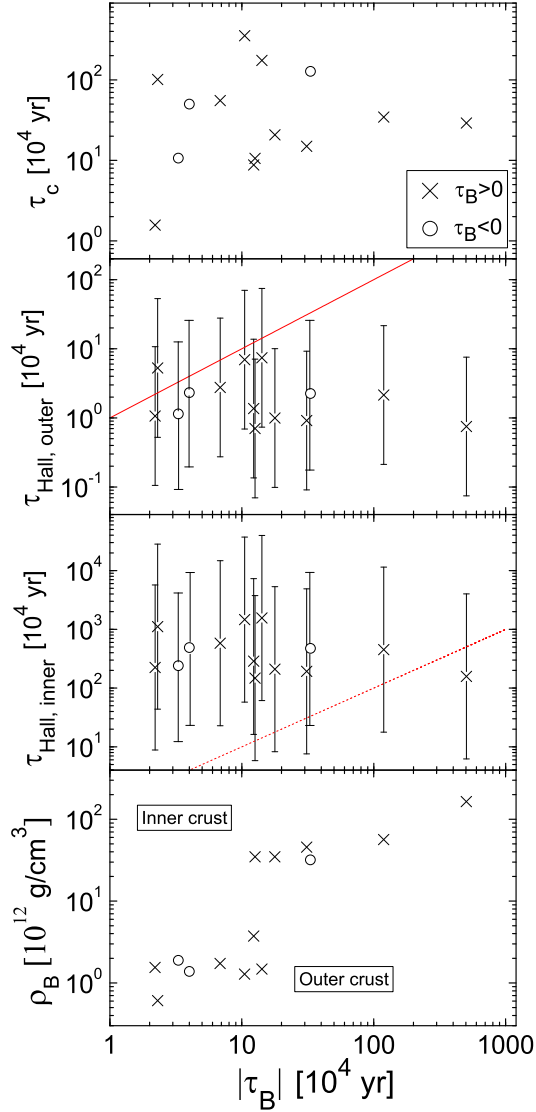}
\caption{\label{Fig:4} $\tau_{\rm c}$, $\tau_{\rm Hall,~outer}$, $\tau_{\rm Hall,~inner}$ and $\rho_{\rm B}$ versus $|\tau_{\rm B}|$. In all panels: the crosses and circles represent $\tau_{\rm B}>0$ and $\tau_{\rm B}<0$, respectively. In the second and third panels: the crosses, circles, solid line and dotted line represent $\tau_{\rm Hall,~outer}(\rho_{12}=1)$, $\tau_{\rm Hall,~inner}(\rho_{14}=1)$, $\tau_{\rm Hall,~outer}=\tau_{\rm B}$ and $\tau_{\rm Hall,~inner}=\tau_{\rm B}$, respectively. In bottom panel: The outer crust and inner crust mean the density values that are derived from the equations $\tau_{\rm Hall,~outer}=\tau_{\rm B}$, and $\tau_{\rm Hall,~inner}=\tau_{\rm B}$, respectively.}
\end{figure}

In the top panel of Figure \ref{Fig:4}, it is shown that there is no significant correlation between $\tau_{\rm c}$ and $|\tau_{\rm B}|$. In the second and third panels, we show that $\tau_{\rm
Hall,~Outer}$ is consistent with $|\tau_{\rm B}|$ for eight pulsars and $\tau_{\rm
Hall,~Inner}$ is for the other six pulsars. Here, the outer and inner crusts are defined by $0.25<\rho_{12}<4$ and $0.25<\rho_{14}<4$, respectively, yielding the upper and lower limits for $\tau_{\rm Hall}$ plotted in the second and third panels of Figure \ref{Fig:4}. This implies that the Hall drift is responsible for
their observed long-term evolutions in $B$. In other words,
the evolution of their crustal magnetic fields, rather than that in their cores, dominates the observed long term spin-down evolution of these young pulsars. Assuming $\tau_{\rm Hall}=|\tau_{\rm B}|$, we obtain an effective density $\rho_{\rm B}$, which may be useful to indicate the location of the majority of the magnetic field lines in the NS crusts. The values of $\rho_{\rm B}$ are shown in the bottom panel of Figure \ref{Fig:4}.

The Hall drift can also pump energy from an internal strong toroidal
field to the dipolar poloidal component on a timescale of $\tau_{\rm
Hall}$ (Pons et al. 2012; Gourgouliatos \& Cumming 2014), resulting in increased $B$ for a pulsar. Therefore the Hall drift can cause both long term decrease and increase of $B$ in a pulsar. Figure 3 shows that all the fourteen pulsars are quite young (with $10^4<\tau_{\rm c}<3\times 10^6$~yr) and only three of the four pulsars are observed with long term increase of $B$, i.e., $\tau_B<0$. This is consistent with our previous result that
 the long-term decrease of magnetic fields are more
frequently observed for young pulsars (Paper~I), suggesting that the Hall drift tends to generate more
magnetic field decay than increase.

Very recently, Geppert et al. (2013) found that the Hall drift can
produce small-scale strong surface magnetic field anomalies (spots)
due to the interaction of the ``initially" dipolar field with the
strong toroidal crustal component, which is fundamental for
generating observable radio emission. Besides, there is a large
scale dipolar poloidal component which has impact on the braking
behavior, and its evolution due to Hall drift may be responsible for
the long-term monotonic evolutions of $\dot\nu$.

\subsection{Short-term magnetic field oscillations}

The above discussed theoretical expectation that the Hall drift can produce both increase and decrease of $B$ suggests that an oscillatory mode of $B$ in a NS crust may also be produced; however, the Hall timescales (either $\tau_{\rm Hall,~Outer}$ or $\tau_{\rm Hall,~inner}$) are too long for the observed periodicities in pulsar timing noises (with $0.4<T<4.3$~yr as listed in Table 1). On the other hand, the diffusive motion of the magnetic fields can perturb the background
magnetic fields (here the dipole magnetic fields) at the base of the
NS crust. Such perturbations propagate as circularly polarized
``Hall waves" along the background field lines upward into the lower
density regions in the crusts (Thompson \& Duncan 1996). The Hall
waves can strain the crust with a wave period (C2004),
\begin{equation}\label{WaveP}
P_{\rm Hall}\simeq \frac{\tau_{\rm Hall,b}}{n^2}\simeq \frac{10^7
{\rm yr}}{B_{12}n^2},
\end{equation}
where $\tau_{\rm Hall,b}$ is the Hall timescale at the base of the
crust, and $n$ is the number of nodes over the crust. The elastic
response of the crust to the Hall wave can induce angular
displacement $\theta_{\rm s}=\frac{d\xi}{dz}=-\frac{B_{\rm
z}}{4\pi\mu}\delta B$, where $\xi$ is the fluid displacement, $z$ is
the depth of the crust, and $\mu$ is the shear modulus. The maximum
strain occurs at or above the turning point. Using the WKB scaling,
it is found that at the turning point (C2004),
\begin{equation}\label{Strain}
\theta_{\rm s,turn}=3\times 10^{-7} B_{12}^2n^{13/9}\frac{\delta
B_{\rm b}}{B},
\end{equation}
in which $\delta B_{\rm b}$ is the amplitude of the mode at the base
of the crust. The density at the turning point can then be
calculated from (C2004)
\begin{equation}\label{rho}
\rho_{\rm turn}\simeq \frac{2\times 10^{12}~{\rm g~cm^{-3}}}{Y_e
n^{4/3}}.
\end{equation}

Since most of the dipole field lines are collected at the polar
regions of NSs, the strains at the regions are especially strong.
The elastic responses of the crusts to the waves induce changes of
the areas of polar regions, producing both pulse shape changes and
oscillations in $\dot\nu$. The quasi-periodical oscillations in
$\dot\nu$ and time residuals can be caused by the Hall waves, i.e.,
$P_{\rm Hall}\sim T$. The calculated parameters from the
above equations are also listed in Table 1. The values of
$\rho_{\rm turn}\thicksim 10^8~{\rm g/cm^3}$ are moderate for the
theoretically predicted density at the turning point. Above the
turning point, the perturbation on magnetic fields $\delta B$
decreases to match the boundary condition $\delta B=0$. The
variation in $\dot \nu$ roughly correlates with $\theta_{\rm
s,turn}$ as $\sin^2\theta_{\rm s,turn}/\sin^2\theta\simeq\Delta\dot
\nu_{0}/\dot \nu_0$, where $\Delta\dot \nu_{0}$ is the amplitude of
the oscillations in $\nu_{0}$.

From Equation~(\ref{vdot}), we have $\Delta\dot \nu_0/\dot
\nu_0=4\kappa$. For B1828$-$11, $\Delta\dot \nu_0/\dot \nu_0=4\times
10^{-3}$, which is roughly the same as $0.5\Delta\dot \nu/\dot
\nu=0.36$ (listed in Table 1); here $\Delta\dot \nu$ is defined as
the peak-to-peak amplitude of the oscillations. The small difference
between $\Delta\dot \nu_0/\dot \nu_0$ and $0.5\Delta\dot \nu/\dot
\nu$ is due to the partial cancellation of the two harmonics with
the same amplitude, but different phases. For the other thirteen pulsars,
$\kappa$ is not available and we thus assume $\Delta\dot \nu_0/\dot
\nu_0=0.5\Delta\dot \nu/\dot \nu$, which is reasonable because only
one significant periodicity has been observed for each of them.
Assuming $\theta=\pi/4$, $\theta_{\rm s,turn}$ can be calculated, as listed in Table 1. Finally we can constrain
$\delta B_{\rm b}/B$ (the last column in the two tables), which is
consistent with $\delta B_{\rm b}/B\sim 1$ usually assumed in
theoretical calculations (e.g. Thompson \& Duncan 1996).

\section{Summary, conclusion, discussion and future perspectives}

In this work, we applied the phenomenological model of pulsar timing noise we developed recently to the observed precise timing data of young radio pulsars, and analyzed the influence of the Hall effects in NS crusts on the magnetic field evolutions. This is the first time that the two aspects of the Hall effect, i.e., Hall drift and waves, are correlated with observational data for individual pulsars. We obtained the following conclusions:
\begin{enumerate}
\item The $\dot\nu$ evolutions for most of the pulsars in L2010 sample can be described adequately with our phenomenological model consisting a long-term monotonic change and short-term oscillations in $\dot\nu(t)$;
\item The observed long-term monotonic changes in $\dot\nu(t)$ can be interpreted with the Hall drifts in NS crusts;
\item The observed short-term oscillations in $\dot\nu(t)$ and in pulse shapes can be produced by the Hall waves in NS crusts.
\end{enumerate}

In a unified view, the long-term evolutions in $\dot\nu$ can be
understood as that they are caused by the Hall waves with very low
harmonics ($n\lesssim10$). Actually, Goldreich \& Reisenegger (1992)
conjectured a turbulent Hall cascade, transferring energy from large
to small scales, with an energy spectrum $E_k\varpropto k^{-2}$
(where $k$ is the wavenumber). However, it is still unclear that why the high harmonics (e.g. $n\sim1000$) dominate in all pulsars we studied here. It is probably due to some selection biases
that the observational time spans for all pulsars are only a few
decades. Indeed, Hobbs et al. (2010) found that the structures seen
in the timing noise vary with data span; as more data are collected,
more quasi-periodic features are observed. To ultimately address
these questions, the power spectrum and the physical origin of the
Hall waves need to be investigated in detail in future works.

As we have discussed above and shown previously, the long-term evolution of $B$ of young pulsars are quite similar to that of the (young) pulsars studied here. In addition, the huge spans of the reported braking indices of all pulsars can be naturally explained by short-term oscillations of their $B$ with characteristics also similar to that of these pulsars. Therefore it is reasonable to assume that the Hall effects we studied here are also responsible for the observed spin evolutions of essentially all pulsars, including millisecond radio pulsars. Physically both a long-lived core based dipolar field and a crustal field should coexist and be superimposed for a young pulsar. However, it is likely that the ``observed" features, with e.g. $10^4<\tau_B< 10^7$~yr, are caused by the crustal field, whose evolution proceeds faster than that of the core field, due to Hall effect and finite conductivity in its crust. As the pulsar ages, its crustal field is gradually dissipated and the core field begins to dominate its spin-down evolution. In this scenario we would expect much longer $\tau_B$ for millisecond pulsars.

Unfortunately, currently the data on the evolutions of $\dot\nu$ of millisecond
pulsars are not available with comparable details as for those
normal radio pulsars studied here (L2010). This is because that
millisecond pulsars have much weaker dipole magnetic fields,
implying much smaller $\theta_{\rm s,turn}$ and
$\Delta\dot\nu_{0}/\dot \nu_0$, which is actually the basis for
using them to detect gravitational waves. Availability of higher
quality timing data of millisecond pulsars from the current on-going
pulsar timing array observations (Haasteren et al. 2011; Demorest et
al. 2013; Manchester et al. 2013) may allow both the long-term
monotonic evolutions and short oscillations to be identified for
individual millisecond pulsars, which will improve the current
gravitational-wave limits (Shannon et al. 2013).

We thank J. Y. Liao, J. Xu and P. F. Wang for useful discussions.
SNZ acknowledges partial funding support by 973 Program of China under grant 2009CB824800,
by the National Natural Science Foundation of China under grant Nos. 11133002, 11373036
and 10725313, and by the Qianren start-up grant 292012312D1117210.

\begin{table}[h]
\caption{Observed and derived parameters for the pulsars with
long-term monotonic changes and short-term quasi-periodical
oscillations in $\dot\nu$; the former are taken from L2010.}

\centering

\begin{tabular}{c|cccccccccccccc}

\hline \hline

Pulsar & $\nu_0$ & $-\dot\nu_0$ & $\ddot\nu_{\rm L}$ & B & $\tau_{\rm B}$ & $T$ & $\frac{\Delta\dot\nu}{\dot\nu}$ & $n$ & $\theta_{\rm s,turn}$ & $\frac{\delta B_{\rm b}}{B}$\\
& s$^{-1}$ & 10$^{-14}$s$^{-2}$&10$^{-25}$~s$^{-3}$& 10$^{12}$~G & 10$^{4}$~yr & yr & \% & 10$^3$ & 10$^{-2}$ & unity \\
\hline
B0740$-$28  & 5.99 & 604.4 & 5.68  & 5.36 & 2.2  & 0.4 & 0.66  & 2.28 & 4.06  & 0.07   \\
B0919+06    & 2.32 & 7.34  & 1.23  & 2.45 & -4.0 & 1.6 & 0.68  & 1.6  & 4.1   & 0.54   \\
B0950+08    & 3.95 & 3.59  & -0.15 & 0.77 & 14.2 & 1.2 & 0.84  & 3.33 & 4.58  & 2.10   \\
B1540$-$06  & 1.41 & 1.75  & 0.04  & 2.53 & -33.1& 4.3 & 1.71  & 0.96 & 6.54  & 1.68   \\
B1642$-$03  & 2.58 & 11.84 & 0.10  & 2.66 & 119  & 4.3 & 2.52  & 0.94 & 7.96  & 1.91  \\
B1714$-$34  & 1.52 & 22.75 & -0.13 & 8.12 & 12.6 & 4.3 & 0.79  & 0.54 & 4.45  & 0.26   \\
B1818$-$04  & 1.67 & 17.70 & 0.20  & 6.23 & 31.0 & 1.2 & 0.85  & 1.17 & 4.61  & 0.15   \\
B1826$-$17  & 3.26 & 58.85 & 0.16  & 4.18 & 12.3 & 2.2 & 0.68  & 1.06 & 4.12  & 0.34  \\
B1828$-$11  & 2.47 & 36.7  & 8.72  & 4.99 & -3.3 & 1.4 & 0.71  & 1.2  & 4.2   & 0.20   \\
B1839+09    & 2.62 & 7.50  & -0.63 & 2.07 & 6.85 & 1.1 & 2.00  & 2.12 & 7.07  & 0.86   \\
B1903+07    & 1.54 & 11.76 & -0.15 & 5.73 & 17.8 & 1.4 & 6.80  & 1.10 & 13.1  & 0.53  \\
B1907+00    & 0.98 & 5.33  & 0.08  & 7.59 & 504  & 1.4 & 0.75  & 0.96 & 4.33  & 0.12   \\
B1929+20    & 3.74 & 5.86  & -1.57 & 1.08 & 2.3  & 1.7 & 0.31  & 2.3  & 2.8   & 1.10   \\
B2148+63    & 2.63 & 1.18  & -0.07 & 0.82 & 10.5 & 2.6 & 1.69  & 2.18 & 6.50  & 4.90  \\
\hline \hline
\end{tabular}

\end{table}

\end{document}